\documentclass[10pt,conference]{IEEEtran}

\usepackage[export]{adjustbox}
\usepackage{amsmath}
\usepackage{amssymb}
\usepackage{booktabs}
\usepackage{caption}
\usepackage{datetime}
\usepackage{hyperref}
\usepackage{multirow}
\usepackage{subcaption}
\usepackage{float}
\usepackage[]{cite}
\usepackage[most]{tcolorbox}
\usepackage{tikz}

\usetikzlibrary{arrows.meta, positioning, shapes}

\tikzset{
  every node/.style={draw, ellipse, minimum height=2cm, minimum width=4cm, align=center},
  blackarrow/.style={-{Latex}, line width=1.1pt},
  grayarrow/.style={-{Latex}, line width=1pt, black!20},
  orangearrow/.style={-{Latex}, line width=2pt, orange},
  lightorangearrow/.style={-{Latex}, line width=2pt, orange!30}
}

\title{It's Not Easy Being Green: On the Energy Efficiency of Programming Languages}

\makeatletter
\newcommand{\linebreakand}{%
  \end{@IEEEauthorhalign}%
  \hfill\mbox{}\par%
  \mbox{}\hfill\begin{@IEEEauthorhalign}%
}
\makeatother

\author{
  \IEEEauthorblockN{Nicolas van Kempen}
  \IEEEauthorblockA{
    University of Massachusetts Amherst, USA\\
    \texttt{nvankempen@cs.umass.edu}
  }
\and
  \IEEEauthorblockN{Hyuk-Je Kwon\IEEEauthorrefmark{1}}
  \IEEEauthorblockA{
    Unaffiliated, USA\\
    \texttt{hyukjekwon@gmail.com}
  }
\and
  \IEEEauthorblockN{Dung Tuan Nguyen\IEEEauthorrefmark{1}}
  \IEEEauthorblockA{
    Eureka Robotics, Vietnam\\
    \texttt{ntddebugger@gmail.com}
  }
\linebreakand
  \IEEEauthorblockN{Emery D. Berger\IEEEauthorrefmark{1}}
  \IEEEauthorblockA{
    University of Massachusetts Amherst, USA\\
    Amazon Web Services, USA\\
    \texttt{emery@cs.umass.edu}
  }
}

\begin{document}

\renewcommand*{\thefootnote}{\fnsymbol{footnote}}
\maketitle
\footnotetext[1]{Work done at the University of Massachusetts Amherst.}
\renewcommand*{\thefootnote}{\arabic{footnote}}

\begin{abstract}
    Does the choice of programming language affect energy consumption? Previous
highly visible studies have established associations between certain programming
languages and energy consumption.
A causal misinterpretation of this work has
led academics and industry leaders to use or support certain languages based on
their claimed impact on energy consumption. This paper tackles this causal question directly:
it develops a detailed
causal model capturing the complex relationship between programming language
choice and energy consumption. This model identifies and incorporates several
critical but previously overlooked factors that affect energy usage. These
factors, such as distinguishing programming languages from their
implementations, the impact of the application implementations themselves, the number of active cores,
and memory activity, can significantly skew energy consumption measurements if not
accounted for. We show---via empirical experiments, improved
methodology, and careful examination of anomalies---that when these factors are
controlled for, notable discrepancies in prior work vanish.
Our analysis suggests that the choice of programming language
implementation has no significant impact on energy consumption beyond execution time.

\end{abstract}

\section{Introduction}

\newcommand{\carbdiox}{CO\textsubscript{2}}

The acceleration of climate change due to use of fossil fuels has
driven an increased focus on efforts to decrease both the energy consumption and carbon footprint of computer systems~\cite{10.1145/3620666.3651374,
10.1145/3652963.3655048, radovanovic2021carbonawarecomputingdatacenters}.
In 2018, an estimated 1\% of total global energy consumption
was attributed to datacenters alone~\cite{doi:10.1126/science.aba3758}.
Modern machine learning workloads---especially training---can generate hundreds of tons of \carbdiox{}
emissions~\cite{DBLP:conf/acl/StrubellGM19, DBLP:journals/corr/abs-2104-10350}.
For example, Meta reports that training the Llama2 large language model generated an estimated 539 tons of \carbdiox{}
emissions~\cite{touvron2023llama2openfoundation}.

Programmers care about energy when building applications, but often lack tools to effectively address this concern~\cite{DBLP:conf/icse/ManotasBZSJSPC16}.
According to an influential line of work, one potential way to reduce
energy consumption is to choose a different programming language. This
work analyzes a wide selection of programming languages and workloads,
and concludes that different programming languages consume widely
varying amounts of energy~\cite{10.1145/3125374.3125382,
  10.1145/3136014.3136031, DBLP:journals/scp/PereiraCRRCFS21}. The
centerpiece of these papers is a ranking of
programming languages by energy efficiency (reproduced in part in Table~\ref{table:paraphrase}).

\begin{figure}[tp]
    \centering
    \begin{subfigure}{\linewidth}
        \centering
        \includegraphics[width=0.7\columnwidth]{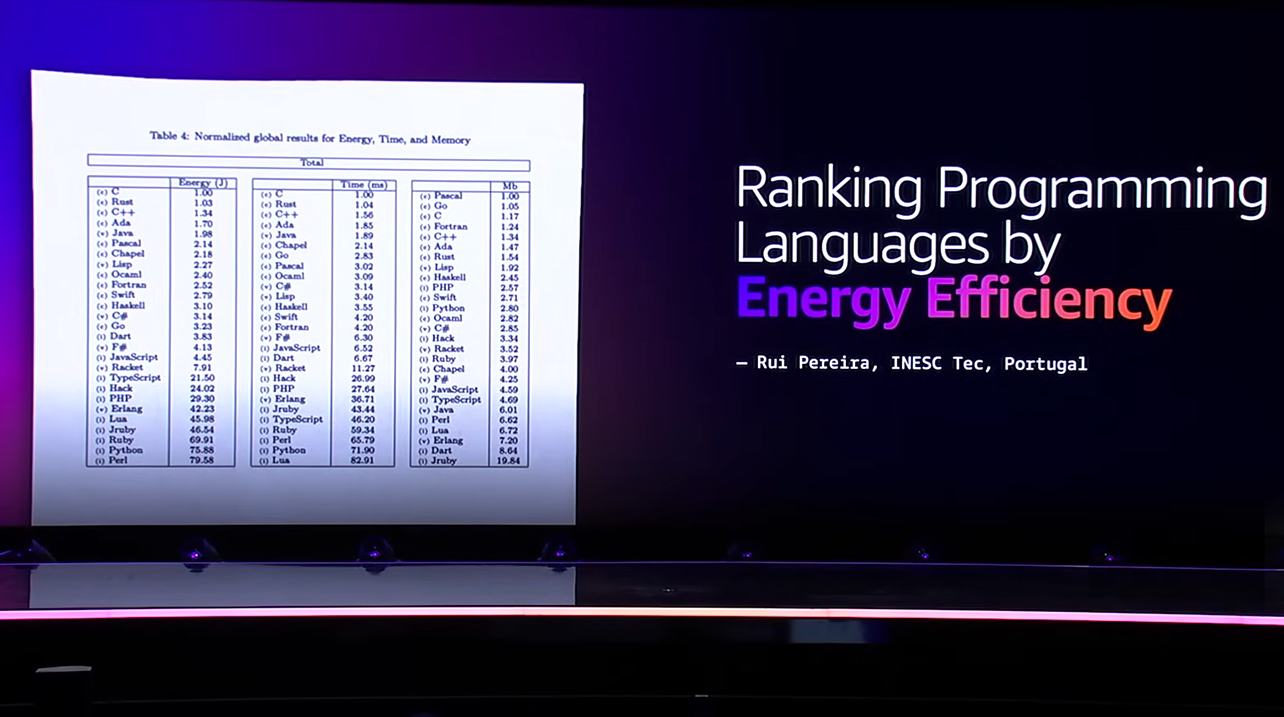}
        \caption{Keynote talk at AWS re:Invent 2023:
            ``There is no reason why you should not be programming in Rust,
            if you are considering cost and sustainability to be high priorities''.\\}
    \end{subfigure}
    \begin{subfigure}{\linewidth}
        \centering
        \includegraphics[width=0.7\columnwidth]{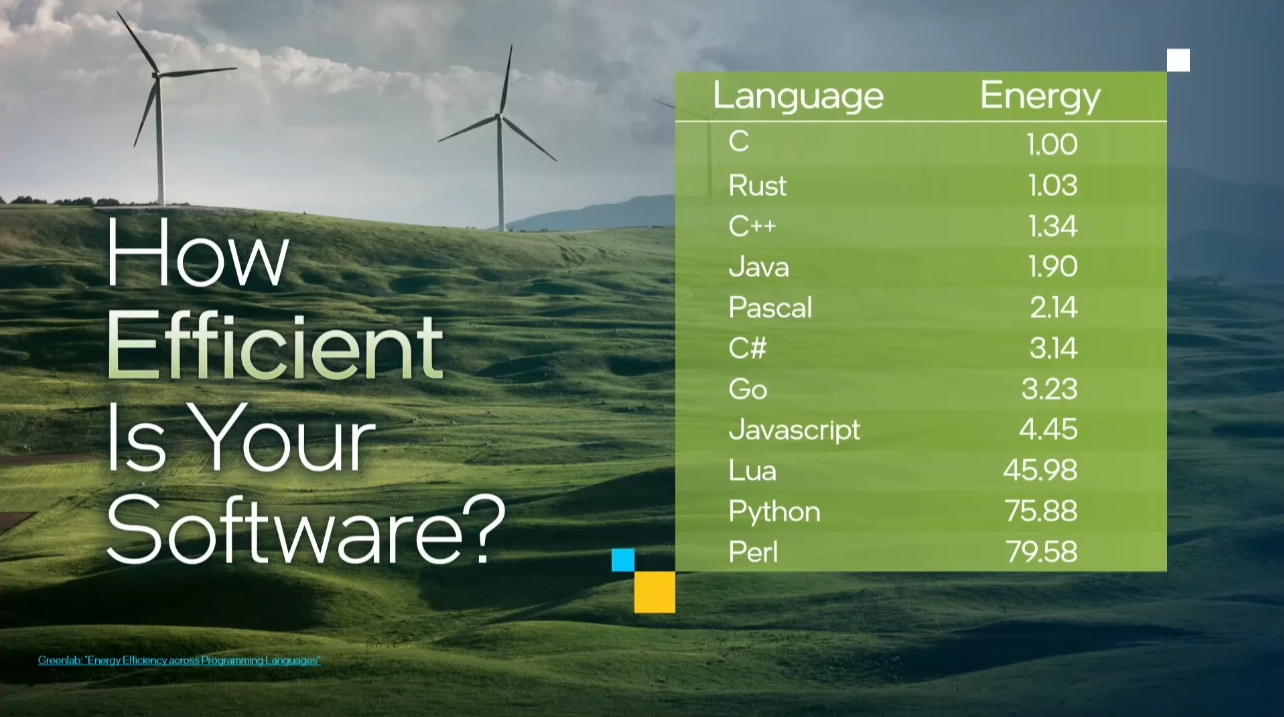}
        \caption{Keynote talk at
        KubeCon + CloudNativeCon Europe 2022: ``Coding with Python over Rust for
        large scale applications can mean a difference of up to $75\times$ in energy usage''.}
    \end{subfigure}
    \caption{\textbf{Industry keynotes advocating for Rust over other languages due to reduced energy consumption, based on the results of Pereira et al.~\cite{10.1145/3125374.3125382,
  10.1145/3136014.3136031, DBLP:journals/scp/PereiraCRRCFS21} (\textsection\ref{section:prior-overview}).}
      }
    \label{fig:screenshots}
\end{figure}

These studies have received wide attention, both in academic and industrial circles.
They have been collectively cited over 800 times per Google Scholar (as of October 2025).
The results---especially the ranking of programming languages---have had an unusually visible impact in industry,
and are routinely quoted on social networks and in blog posts.
As Figure~\ref{fig:screenshots} illustrates, the rankings have been cited by
executives and engineers from Amazon~\cite{AmazonSustainability1,
AmazonSustainability2, AmazonSustainability3, AmazonSustainability4}, Intel~\cite{IntelSustainability},
SAP~\cite{SAPSustainability}, and other companies~\cite{GFTSustainability} to argue for business
decisions and to advocate for a shift in programming languages with an eye towards sustainability.
Often, these rankings have been harnessed to support the adoption of Rust, which ranks as one of
the most energy-efficient languages while providing safety guarantees that languages like C and C++ lack.

Despite the fact that these studies are statistical and only establish associations, they have nonetheless been
broadly interpreted as establishing a \emph{causal} relationship, that the choice of programming
language has a direct effect on a system's energy consumption.
This misinterpretation
stems in part from the work's presentation, not only in ranking of languages by efficiency, but also from the specific claim that ``it is
almost always possible to choose the best language'' when considering
execution time and energy consumption~\cite[\textsection{}3.3]{DBLP:journals/scp/PereiraCRRCFS21}.

\begin{figure*}[tp]
    \centering
    \resizebox{0.94\textwidth}{!}{%
      \begin{tikzpicture}
      \node at (0, 0) (lang) {Programming\\Language};
      \node at (4, 2.5) (impl) {Programming\\Language\\Implementation};
      \node at (4, -2.5) (app) {Application\\Implementation};
      \node at (9.5, 6) (cores) {Number of\\Active Cores};
      \node at (9.5, 2.5) (mem) {Memory\\Activity};
      \node at (15, 2.5) (power) {Power};
      \node at (15, -2.5) (time) {Time};
      \node at (19, 0) (energy) {Energy};
      
      \draw[blackarrow] (lang) -- (impl);
      \draw[blackarrow] (lang) -- (app);
      \draw[blackarrow] (impl) -- (mem);
      \draw[blackarrow] (impl) -- (time);
      \draw[blackarrow] (app) -- (mem);
      \draw[blackarrow] (app) -- (time);
      \draw[blackarrow] (app) -- (cores);
      \draw[blackarrow] (cores) -- (power);
      \draw[blackarrow] (mem) -- (time);
      \draw[grayarrow] (mem) -- (power);
      \draw[blackarrow] (power) -- (energy);
      \draw[blackarrow] (time) -- (energy);
      \draw[grayarrow] (impl) -- (cores);
      \end{tikzpicture}
  }
    \caption{
        \textbf{The causal model (represented as a causal diagram) of the relationship between programming language and
          energy consumption presented in this paper (\textsection\ref{section:model})}.
        Gray arrows represent
        comparatively weaker relationships, as Section~\ref{section:model} details:
        programming language implementation has only a minor effect on parallelism,
        and memory activity plays a minimal, less controllable role in energy consumption
        compared to CPU activity.}
    \label{fig:final-model}
\end{figure*}

The above analysis and approach suffer from numerous other
methodological flaws (\textsection\ref{sec:critique},
\textsection\ref{section:methodology}, \textsection\ref{section:eval-examination}). However, the primary focus of
this paper is carefully addressing the question: \emph{\textbf{does the choice
  of programming language affect energy consumption?}}
We embrace this question by
developing a rich causal model of the relationship between programming
language and energy consumption. Figure~\ref{fig:final-model}
presents a causal diagram representing our final model.
Causal diagrams illustrate how different factors influence each other, with arrows showing the direction of these influences~\cite{pearl2009causality}.
We
build this model incrementally, incorporating the various
factors at play and their relationships, and provide quantitative and/or
qualitative evidence supporting each step.

Our model specifically identifies the number of active cores and memory
activity as the key contributors to power draw and consequently energy
consumption.  It highlights the importance of both programming
language and application implementations: language implementation
decisions can add execution time overheads including garbage collection or
just-in-time warmup, while application implementations
dictate the level of parallelism and overall performance of the
program.  These varying implementation
characteristics are the primary drivers of the anomalies we identify in
prior work.

Our experimental approach incorporates several methodological
improvements, correcting technical errors in prior work that led to
negative or otherwise incorrect energy readings.  We present an
improved methodology based on hardware performance counter data that
accurately quantifies average core usage and memory activity.
When controlling for confounds
like varying CPU utilization, we conclude that energy consumed is
directly proportional to execution time, and independent of the choice of
programming language.
These results simplify the task of programmers who are seeking to achieve higher energy efficiency by focusing their efforts on minimizing execution time.

\section{Prior Work}
\label{section:prior-work}
\subsection{Overview}
\label{section:prior-overview}

The line of work that is the point of departure for this paper explores the relationship between the choice of programming language and energy efficiency~\cite{10.1145/3125374.3125382,
10.1145/3136014.3136031, DBLP:journals/scp/PereiraCRRCFS21}. These papers rank
programming languages based on energy consumption, execution time, and memory usage,
and attempt to find associations between these metrics.
In the remainder of this paper, we refer to these papers collectively as ``Pereira et al.''.

In essence, these studies compare the execution time, energy consumption and
memory usage of benchmark implementations in 27
various programming languages.
Table~\ref{table:paraphrase} presents some of the main results from Pereira et al., limited to the languages discussed in this paper,
with anomalous results highlighted in boldface (Section~\ref{sec:critique} discusses this selection in detail).
Their experiments leverage Intel's Running Average Power Limit (RAPL) interface to measure energy consumption,
and GNU \texttt{time} or Python's \texttt{memory\_profiler} to measure
peak or ``total'' memory usage, respectively.

The programs used for comparison are from the
Computer Language Benchmark Game~\cite{CLBG} (CLBG), a corpus of small benchmark
implementations in various languages (22 to 287 lines of code).
The most recent paper in the series~\cite{DBLP:journals/scp/PereiraCRRCFS21} also
incorporates 9 benchmarks from Rosetta Code, a repository of even smaller and simpler code snippets
such as Fibonacci, Ackermann, or Sieve of Eratosthenes (typically under 50 lines of code).
Because Rosetta Code benchmarks are strictly smaller in scope and size than the CLBG benchmarks, we exclude them from consideration in this paper.

These studies first identify a strong relationship between energy and time. By definition, energy is a linear function of time (energy = power $\times$ time),
hence a strong correlation is to be expected. They next investigate whether a fast language is always more energy efficient,
and claim that this is not the case. They report no significant correlation between peak memory usage and energy consumption,
but a strong correlation when considering total memory usage. Finally, they conclude by presenting a
ranking of programming languages by energy efficiency, execution time, and memory usage, which programmers can use
to select a language for their project.

\begin{table}[tp]
    \caption{\label{table:paraphrase}
        \textbf{Partial results from Pereira et al.\ (\textsection\ref{section:prior-overview})},
        showing average execution time and energy consumption for the
        languages discussed in this paper, normalized to C.
        \textbf{Boldface} denotes anomalous results that Sections~\ref{sec:critique}, \ref{section:eval-examination} address.
    }
    \centering\begin{tabular}{l|r|r}
    \multicolumn{1}{l|}{\textit{Language}} & \textit{Execution Time} & \textit{Energy Consumption}\\
    \hline
        C                   & 1.00           & 1.00\\
        Rust                & 1.04           & 1.03\\
        \textbf{C++}        & \textbf{1.56}  & \textbf{1.34}\\
        \textbf{Java}       & \textbf{1.89}  & \textbf{1.98}\\
        \textbf{Go}         & \textbf{2.83}  & \textbf{3.23}\\
        \textbf{C\#}        & \textbf{3.14}  & \textbf{3.14}\\
        JavaScript          & 6.52           & 4.45\\
        PHP                 & 27.64          & 29.30\\
        \textbf{TypeScript} & \textbf{46.20} & \textbf{21.50}\\
        Python              & 71.90          & 75.88\\
        \textbf{Lua}        & 82.91          & \textbf{45.98}
    \end{tabular}
\end{table}

\subsection{Critique}
\label{sec:critique}

As summarized here and discussed in the following sections, Pereira et al.'s
studies suffer from several flaws, which this paper addresses and corrects.

\textbf{Programming Language versus Implementation:} Programming languages
define the syntax and semantics, but it is their
implementations that primarily influence performance. While some languages have
only a single, widely-used implementation such as Rust or Go, others have multiple implementations, each with their
own performance characteristics. For instance, Pereira et al.\ treat
Ruby and JRuby as different languages, while they are in fact two separate
implementations of the same Ruby language.
The papers use different benchmark implementations to compare these two Ruby implementations, confounding their comparison.

\textbf{Widely Varying Benchmark Implementations:} While benchmark
implementations are claimed to employ the ``exact same algorithm''
across languages~\cite[\textsection{}1]{DBLP:journals/scp/PereiraCRRCFS21},
this is in fact not the case. Benchmark implementations notably have highly varied
levels of parallelism and CPU usage, varying degrees of use of third-party
libraries, and non-uniform use of vector instructions.
These important differences are not properties of the languages themselves, and their effect on performance and energy consumption
must be accounted for.

\textbf{Apparent Anomalies in Results:} Some results reported by Pereira et al.\ are counter-intuitive, and are presented without investigation or explanation. C++ is reported as being 34\% less energy efficient
and 56\% slower than C. Since C++ is approximately a superset of C, and both share the
same compiler, optimizations, and code generation backend, we would expect identical energy consumption and
execution time. Similarly, TypeScript is reported as being
$4.8\times$ less energy efficient and $7.1\times$ slower than JavaScript.
TypeScript is a strict superset of JavaScript: any valid JavaScript program is
a valid TypeScript program. The same Node.js runtime system is used for both languages.
The compilation process for TypeScript may insert or modify code to support older JavaScript standards,
but we do not expect this to result in any significant performance overhead.
Java, C\#, and Go are reported as $1.89\times$, $3.14\times$, and $2.83\times$ slower than C, respectively.
These numbers are
unexpectedly high as these implementations are known to be highly optimized~\cite{DBLP:conf/usenix/LionCSY22}.
We expect low overhead from
garbage collection costs, plus initial just-in-time compilation overhead for Java and C\#.
Lua and TypeScript both stand out as exhibiting significantly lower normalized energy consumption than normalized execution time;
this is explained by the fact that nearly all of the Lua and TypeScript benchmark implementations are sequential,
while the implementations of these benchmarks in other languages are parallelized.
Section~\ref{section:eval-contributors} characterizes the impact of the number of active cores on energy consumption.

\textbf{Inappropriate Memory Metrics:} Peak memory usage and ``total'' memory usage are
used to estimate memory activity. The tools used to gather both metrics are both based on
resident set size (RSS), which is a poor proxy for memory usage and activity~\cite{DBLP:conf/osdi/BergerSP23}.
RSS includes memory that may not be actively used by the program,
and crucially does not take cache activity into account. A well-optimized program can
have a large RSS but excellent cache locality: if the cache can fulfill most memory requests, memory activity will be low.
Benchmark implementation specifics heavily influence peak memory usage, which depends primarily on the choice of data structures used throughout the program.
These differences should not be attributed to the languages themselves.

The total memory usage is measured by using
Python's \texttt{memory\_profiler}, which samples the RSS at a frequency of ten times per second.
This metric is directly proportional to execution time,
which is itself directly proportional to energy consumption.

\textbf{Failure to Control for Language Implementation Characteristics:} Most language implementations have an initial cost to import and set up necessary in-memory data structures.
Language implementations using a just-in-time compiler also require an initial
warmup period~\cite{DBLP:journals/ese/TrainiCPT23, DBLP:journals/pacmpl/BarrettBKMT17}.
These start-up costs are amplified by the benchmarks' short execution times, which in some cases run for less than a fraction of a second.
For example, measuring only the first iteration of a short-lived Java benchmark
may not be indicative of Java's overall performance. Section~\ref{section:eval-examination}
shows that this first iteration can be up to $3\times$ slower than subsequent ones.
Garbage collection also can have a significant impact
on both execution time and memory usage~\cite{10.1145/1094811.1094836}, and can be fine-tuned to obtain better performance.

\section{Research Questions}
\label{section:research-questions}
This paper provides numerous improvements to methodology and results over prior work.
Section~\ref{section:model} first introduces and gradually builds a causal model capturing all factors standing between
choice of programming language and energy consumption. Section~\ref{section:methodology} describes the improved methodology used
in this paper, including correction of technical errors and enhanced data gathering with performance counters.
Finally, using both the causal model and improved methodology, Section~\ref{section:eval} addresses the following research questions:
\begin{itemize}
    \item \textbf{RQ1}: Do some language implementations consume more energy than others?
    \item \textbf{RQ2}: What are the key contributors to power draw standing between choice of programming language and energy consumption?
    \item \textbf{RQ3}: Can observed anomalies in prior work be explained through the lens of our causal model?
\end{itemize}

\section{Causal Model for Energy Consumption in Programming Languages}
\label{section:model}
This section builds piece by piece the causal model presented as a diagram in its final form in Figure~\ref{fig:final-model}.
Causal diagrams illustrate how different factors influence each other, with directed edges indicating cause-to-effect relationships~\cite{pearl2009causality}.
Formally, edges indicate that changes in the origin node impact the probability distribution of its descendants.
Our diagrams additionally use gray arrows to represent comparatively weaker relationships, as detailed in the following sections.
Each subsection below gradually adds nodes and edges to the diagram in order to obtain a rich causal model that captures the essential factors in the relationship
between the choice of programming language and energy consumption. Figures~\ref{fig:diagram-0}, \ref{fig:diagram-1}, \ref{fig:diagram-2}, and
\ref{fig:diagram-3} build upon each other, with new elements in orange. They culminate in the final model represented in Figure~\ref{fig:final-model}.

\subsection{Starting Point}
\label{section:model-starting}

\begin{figure}[tp]
    \centering
    \resizebox{0.8\columnwidth}{!}{%
        \begin{tikzpicture}
        \node at (0, 0) (lang) {Programming\\Language};
        \node at (5.5, 0) (energy) {Energy};

        \draw[blackarrow] (lang) -- (energy);
        \end{tikzpicture}
    }
    \caption{\textbf{The simple model used as a starting point in this paper (\textsection\ref{section:model-starting}).} This simple model captures
        the relationship implied in Pereira et al., namely that the choice of programming
        language has a direct impact on total energy consumption.}
    \label{fig:diagram-0}
\end{figure}
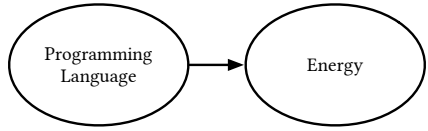

Our starting point, shown in Figure~\ref{fig:diagram-0}, is that the choice of programming
language has a direct effect on energy consumption.
The following sections will explore additional elements that come into play in this relationship,
and build up a complex model exploring the impact of different factors on energy consumption.

\subsection{Programming Languages and Implementations}
\label{section:model-pl-implementation}

\begin{figure}[tp]
    \centering
    \resizebox{\columnwidth}{!}{%
        \begin{tikzpicture}
        \node at (0, 0) (lang) {Programming\\Language};
        \node[draw=orange, line width=1.2pt] at (4, 2) (impl) {Programming\\Language\\Implementation};
        \node at (8, 0) (energy) {Energy};

        \draw[orangearrow] (lang) -- (impl);
        \draw[orangearrow] (impl) -- (energy);
        \end{tikzpicture}
    }
    \caption{\textbf{Programming languages may have multiple implementations (\textsection\ref{section:model-pl-implementation})}, and implementation
        decisions such as garbage collection or using just-in-time compilation have a more direct impact on performance. 
        We are in fact comparing implementations of programming languages, not the languages themselves.}
    \label{fig:diagram-1}
\end{figure}
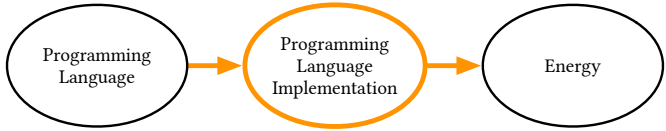

Figure~\ref{fig:diagram-1} introduces the important distinction between the programming languages themselves and their implementations.
Some languages have a single well-known implementation, or a blurry line between language and implementation, such as Rust or Go.
Others have multiple implementations, such as C/C++ with GCC, Clang, or MSVC, Java with various JVMs, Python with CPython and PyPy, or Lua with LuaJIT.
Finally, there can be significant differences between specific \textit{versions} of the same implementation.
Our model treats those different versions as different
implementations altogether. For example, recent CPython releases each successively delivered significant performance improvements,
notably with the 3.11 release achieving a 25\% speedup over 3.10~\cite{python-3.11-performance}.

Of course, programming languages have an effect on implementation possibilities. They may dictate the general memory layout of objects,
the need for a garbage collector, or the possibility of ahead-of-time compilation.
For example, Java has no alternative for memory reclamation other than a garbage collector.
Other dynamic features such as dynamic typing, reflection, or \texttt{eval}
further limit the range of options available to implementers.

The impact of the characteristics of programming languages on their implementations is transitively included in our model: 
the language has a causal influence on the implementation, which in turn has a
direct causal impact on execution time. The performance and energy consumption impact of those language features is mediated by the implementation.
Dynamic dispatch, for example, may be
implemented or optimized with v-tables, fat pointers, monomorphic and polymorphic inline caches, or profile-guided optimization.

\subsection{Decomposing Energy into Power and Time}
\label{section:model-energy-power-time}

\begin{figure}[tp]
    \centering
    \resizebox{\columnwidth}{!}{%
        \begin{tikzpicture}
        \node at (0, 0) (lang) {Programming\\Language};
        \node at (4, 2) (impl) {Programming\\Language\\Implementation};
        \node[draw=orange, line width=1.2pt] at (3, -2) (time) {Time};
        \node[draw=orange, line width=1.2pt] at (8, -2) (power) {Power};
        \node at (5.5, -4.5) (energy) {Energy};
        
        \draw[blackarrow] (lang) -- (impl);
        \draw[lightorangearrow] (impl) -- (power);
        \draw[orangearrow] (impl) -- (time);
        \draw[orangearrow] (time) -- (energy);
        \draw[orangearrow] (power) -- (energy);
        \end{tikzpicture}
    }
    \caption{\textbf{Energy consumption is the product of power and time (\textsection\ref{section:model-energy-power-time}).} Splitting into
        these two components allows separate consideration of each factor.}
    \label{fig:diagram-2}
\end{figure}

Fundamentally, energy consumption is the product of power and time. Figure~\ref{fig:diagram-2} decomposes these two factors.
Section~\ref{section:eval-contributors} demonstrates that power draw is constant across all languages, implementations, and benchmarks. %
Claims of the contrary in prior work were in large part due to missing multicore accounting, directly comparing sequential and concurrent benchmarks.

In theory, runtime systems running in parallel to the user program may affect energy consumption beyond variance in execution time,
since increased core usage will increase power draw, as Section~\ref{section:eval-contributors} details below.
This hidden increased parallelism could occur with a parallel garbage collector or just-in-time compilation threads executing alongside the main program.
In practice, compilation threads are primarily active only during startup, and both are rarely significant compared
to the main program's execution and other general overhead costs.
For this reason, we draw this edge in gray in our model.

\subsection{Contributors to Power Draw}
\label{section:model-power}

\begin{figure}[tp]
    \centering
    \resizebox{\columnwidth}{!}{%
    \begin{tikzpicture}
        \node at (0, 0) (lang) {Programming\\Language};
        \node at (2.5, -2.5) (impl) {Programming\\Language\\Implementation};
        \node at (8, -2.5) (time) {Time};
        \node[draw=orange, line width=1.2pt] at (0, -5) (cores) {Number of\\Active Cores};
        \node[draw=orange, line width=1.2pt] at (5, -5) (mem) {Memory\\Activity};
        \node at (2.5, -7.5) (power) {Power};
        \node at (8, -7.5) (energy) {Energy};
        
        \draw[blackarrow] (lang) -- (impl);
        \draw[blackarrow] (impl) -- (time);
        \draw[blackarrow] (time) -- (energy);
        \draw[blackarrow] (power) -- (energy);
        
        \draw[lightorangearrow] (impl) -- (cores);
        \draw[orangearrow] (impl) -- (mem);
        \draw[orangearrow] (cores) -- (power);
        \draw[lightorangearrow] (mem) -- (power);
        \end{tikzpicture}
    }
    \caption{\textbf{Number of active cores is the primary factor in increased power draw (\textsection\ref{section:model-power}).}
        Memory activity also increases power draw, but to a much lesser extent.
        Figure~\ref{fig:power-memory} quantifies this.}
    \label{fig:diagram-3}
\end{figure}

Previous research has demonstrated that processor and memory energy usage are the
main variable contributors to power draw~\cite{DBLP:conf/micro/Vogelsang10}.
We add these two factors to our model in Figure~\ref{fig:diagram-3}.
For the studied programs on our experimental platform,
the ratio of memory energy consumption to CPU energy consumption remains between 2 and 8\% (Section~\ref{section:eval-contributors}).
Further, memory activity is not something that is easily controlled by the programmer.
While certain algorithms or data structures maximize cache locality,
there is eventually a limit stemming from the LLC's fixed, relatively small size. Programming languages
or their implementations may impose additional overhead with a garbage collector, memory allocation
overhead, or object memory layout. Because of the difficulty of controlling memory activity, as well as its comparatively
smaller impact compared to CPU activity on power draw, we draw this edge in gray in our model.

\subsection{Application Implementations}
\label{section:model-benchmarks}

We add a final node to our diagram: the application implementations themselves, which have the most direct impact on energy consumption.
Figure~\ref{fig:final-model} presents the final causal model developed in this paper.
Application implementation decisions dictate the level of parallelism, the algorithms used, and the data structures employed.

\subsection{Other Factors}

There are many other factors that may affect energy consumption.
CPU voltage and frequency scaling has a direct impact on power consumption, and there is active research focusing on reducing or
increasing frequency based on memory activity or other factors to draw less power for a comparatively smaller performance penalty~\cite{10.1145/581630.581668}.
Specific processor model and architecture also have an impact on power draw, with newer processors generally being more energy-efficient.
These factors are unrelated to choice of programming languages and are controlled for in our experiments.

\section{Enhanced Methodology}
\label{section:methodology}
Our methodology improves upon prior work in several ways. First, it corrects technical errors
in the measurement tool used by Pereira et al.~\cite{gslrepo}. It also enhances data gathering by
collecting performance counter readings, which we use to identify key contributors to energy consumption.
Finally, it expands the set of benchmarks used to include two additional widely-used benchmark suites.

\subsection{Measurement Tool}
\label{section:measurement-tool}

We develop a new measurement tool for this study,
correcting errors found in Pereira et al.'s implementation~\cite{gslrepo},
and adding performance counter readings capturing the number of cores used and memory activity.
The tool reads performance counters at the beginning and end of each benchmark run,
and samples energy consumption readings once per second.
Periodic sampling is necessary to ensure correct energy readings, as the next section explains.

\textbf{RAPL:}
Starting with the Sandy Bridge microarchitecture (2011), most Intel processors provide a power
management interface called Running Average Power Limit (RAPL). RAPL allows
measurement of the energy consumption of various parts of the system
\cite{DBLP:journals/tompecs/KhanHNNO18}. The interface is exposed via
Model Specific Registers (MSRs), of which the following measure energy consumption:
\begin{itemize}
    \item \texttt{MSR\_RAPL\_POWER\_UNIT} contains information used to convert
        the raw energy status counter value to joules.
    \item \texttt{MSR\_PKG\_ENERGY\_STATUS} contains the raw energy status
        counter for the entire processor package (denoted PKG in our graphs).
    \item \texttt{MSR\_DRAM\_ENERGY\_STATUS} contains the raw energy status
        counter for the random access memory attached to that processor.
\end{itemize}

MSRs track energy consumption at the granularity of an entire package. There
is no way to assign energy consumption to a specific thread or process, but only to
the system as a whole.
In particular, there is no way to eliminate overhead from the measurement tool and
other background processes from the measurement. The measurement tool
must therefore be as lightweight as possible. We disable all
non-essential processes, keeping only those necessary for the machine to
operate, and assume these processes remain constant across all
experiments.

Further, RAPL samples include all cores, even if the program under test
only uses a single core. If a benchmark is single-threaded or generally
uses fewer cores than available, idle cores will be included in the
energy consumption measurement. Therefore, using a varying level of parallelism across
benchmark implementations can result in unfair comparison, as idle cores will add some constant
energy consumption to each sample.

An inspection of Pereira et al.'s data files~\cite{gslrepo} reveals the presence of multiple
negative energy readings. Of course, negative energy consumption is not physically possible.
We expect that these values were removed by the outlier removal process described in their work and did not affect final results.
To ensure correct energy readings, we rectify the following errors in the tool used by previous work:
\begin{itemize}
    \item {\textbf{Erroneous inclusion of upper bits of status counters:}} The previous tool incorrectly includes the upper 32 bits of the energy status counters in their accounting.
        These upper bits are reserved and must be discarded;
        our tool tracks only the bottom 32 bits, which contain the actual count of consumed energy.
    \item {\textbf{Failure to prevent overflow of status counters:}} The previous tool reads only a start and end value of the energy status counter, but that counter overflows after
        ``around 60 seconds when power consumption is high''~\cite{IntelManual}.
        To avoid overflow for long-running computations, our tool uses a separate thread that periodically (at 1Hz) reads the contents of this
        register.
       
    \item {\textbf{Erroneous floating point arithmetic:}} Finally, the previous tool incorrectly subtracts counter readings after scaling and
        conversion to floating point. This conversion will result in negative
        or otherwise incorrect results when overflow occurs. Our tool avoids this problem by
        converting results to floating point only after subtracting consecutive readings.
\end{itemize}

\textbf{Performance Counters:}
Hardware performance counters are a feature of modern processors that track
various events, such as total cycles, cache misses, branch misses, and so on.
They are often used to analyze program performance, notably when profiling.
These counters are 64 bits in length and thus not susceptible to overflow, so it
suffices to read them once at the beginning and once at the end of each experiment.

\textit{Memory Activity:}
Section~\ref{section:model-power} highlights the impact of memory activity on energy consumption:
a majority of memory energy consumption is dependent on the rate of read and write operations.
We monitor the last level cache (LLC) using performance counters to estimate memory activity.
Each LLC hit
means that the cache already had a copy of the data, and therefore no further memory activity occurs.
On the other hand, each LLC miss means that the system must fetch data from memory.
Therefore, the number of LLC misses is a good proxy for memory activity.

\textit{Average CPU Usage:}
There are multiple ways to measure average core usage while a program is running.
For simplicity, we leverage performance counters provided by the operating system. The Linux kernel
exposes the \texttt{task-clock} software event counter, which aggregates in nanoseconds
the time spent on all processor cores. To obtain average core usage,
we divide this counter's value by the total execution time of the program.

\subsection{Benchmarks}
\label{section:benchmarks}

This paper extends the evaluation performed by the original Pereira et al.\ study by incorporating two additional large-scale benchmark suites.
We first include all possible benchmark implementations from the Computer Language Benchmark Game used by Pereira et al.\ that successfully compile and run (118 benchmarks).
Using the same set of benchmarks allows reproduction and re-examination of results in Section~\ref{section:eval-examination}.
In addition, the analysis in Sections~\ref{section:eval-energy} and~\ref{section:eval-contributors} includes two more widely-used benchmark suites:
SPEC CPU 2017~\cite{speccpu2017} (18 benchmarks) and DaCapo Chopin~\cite{10.1145/3669940.3707217} (22 benchmarks).
The following subsections provide a brief overview of each of these benchmark suites.

\begin{table}[tp]
    \caption{
        \textbf{Computer Language Benchmark Game benchmarks (\textsection\ref{section:benchmarks})},
        along with the range of lines of code (LoC) across languages for each benchmark as reported by \texttt{cloc}.
    }
    \label{table:clbg-descriptions}
    \centering\begin{tabular}{l | l | l}
        \multicolumn{1}{l|}{\textit{Benchmark}} & \multicolumn{1}{l|}{\textit{LoC}} & \multicolumn{1}{l}{\textit{Description}}\\ \hline
        binary-trees       & 36--98  & Memory allocator stresstest\\
        fannkuch-redux     & 40--158 & Permutation algorithm\\
        fasta              & 84--287 & Random DNA generation\\
        k-nucleotide       & 57--202 & Frequency counting\\
        mandelbrot         & 32--140 & Fractal rendering\\
        n-body             & 78--157 & Physics simulation\\
        pidigits           & 41--149 & Arbitrary precision\\
        regex-redux        & 22--111 & String processing\\
        reverse-complement & 34--257 & String transformation\\
        spectral-norm      & 33--156 & Linear algebra
    \end{tabular}
\end{table}

\textbf{Computer Language Benchmark Game:}
The Computer Language Benchmark Game~\cite{CLBG} (CLBG) is a collection of small programs implemented in many different programming languages
(22 to 287 lines of code).
We use the same benchmark implementations as Pereira et al.~\cite{gslrepo},
which is in effect a snapshot of the fastest versions of the benchmarks as available on
the CLBG repository at the time of the original work. Table~\ref{table:clbg-descriptions}
provides a brief description of each benchmark. We limit our analysis to eleven
languages across thirteen implementations: C, C\#, C++, Go, Java,
JavaScript, Lua (Lua and LuaJIT), PHP, Python (CPython and PyPy), Rust, and TypeScript. These languages comprise 11 of the
top 12 most popular general programming languages in
the 2024 Stack Overflow survey~\cite{StackOverflowSurvey} (excluding Kotlin).

We attempt to compile and run all benchmark implementations with sources present in the repository,
and omit those without source code or which fail with compilation or run-time errors.
After these omissions, the analysis below spans 118 out of 130 possible language implementation / benchmark pairs.
Some benchmarks
make use of third-party libraries: we use the default package manager's version
when available, or manually upgrade to the latest available version.

We stress that while the CLBG benchmarks are not
necessarily representative of real-world applications, the causal
analysis this paper develops is largely independent of the details of
the benchmark implementations.  It instead highlights the impact of
high-level \emph{properties} of the benchmark implementations, such as their
degree of parallelism and cache activity.

\begin{table}[tp]
    \caption{
        \textbf{SPEC CPU 2017 benchmarks (\textsection\ref{section:benchmarks})}.
    }
    \label{table:spec-descriptions}
    \centering\begin{tabular}{l | l | l}
        \multicolumn{1}{l|}{\textit{Benchmark}} & \multicolumn{1}{l|}{\textit{Language(s)}} & \multicolumn{1}{l}{\textit{Description}}\\\hline
        perlbench & C       & Perl interpreter\\
        gcc       & C       & GNU C compiler\\
        mcf       & C       & Route planning\\
        omnetpp   & C++     & Discrete-event simulation\\
        xalancbmk & C++     & XML to HTML conversion\\
        x264      & C       & Video compression\\
        deepsjeng & C++     & Alpha-beta tree search\\
        leela     & C++     & Monte Carlo tree search\\
        exchange2 & Fortran & Recursive solution generator\\
        xz        & C       & General data compression\\
        \hline
        bwaves    & Fortran         & Explosion modeling\\
        cactuBSSN & C++, C, Fortran & Physics: relativity\\
        lbm       & C               & Fluid dynamics\\
        wrf       & Fortran, C      & Weather forecasting\\
        cam4      & Fortran, C      & Atmosphere modeling\\
        pop2      & Fortran, C      & Wide-scale ocean modeling\\
        imagick   & C               & Image manipulation\\
        nab       & C               & Molecular dynamics\\
        fotonik3d & Fortran         & Computational electromagnetics\\
        roms      & Fortran         & Regional ocean modeling
    \end{tabular}
\end{table}

\begin{table}[tp]
    \caption{
        \textbf{DaCapo benchmarks (\textsection\ref{section:benchmarks})}.
    }
    \label{table:dacapo-descriptions}
    \centering\begin{tabular}{l | l}
        \multicolumn{1}{l|}{\textit{Benchmark}} & \multicolumn{1}{l}{\textit{Description}}\\\hline
        avrora     & AVR controller simulation\\
        batik      & Render SVG files\\
        biojava    & Protein sequence analysis\\
        cassandra  & NoSQL database\\
        eclipse    & IDE workload\\
        fop        & Render PDF files\\
        graphchi   & ALS matrix factorization\\
        h2         & SQL database\\
        h2o        & Machine learning\\
        jme        & Render video game frames\\
        jython     & Python implementation\\
        kafka      & Stream processing\\
        luindex    & Text indexing\\
        lusearch   & Text search\\
        pmd        & Static analysis\\
        spring     & Web framework\\
        sunflow    & Render photorealistic images\\
        tomcat     & Web server\\
        tradebeans & Stock market simulation\\
        tradesoap  & Stock market simulation\\
        xalan      & XML transformation\\
        zxing      & Barcode reader
    \end{tabular}
\end{table}

\textbf{SPEC CPU 2017:}
SPEC CPU 2017~\cite{speccpu2017} is a widely used benchmark suite for evaluating systems on C, C++ and Fortran code.
The suite consists of large real-world programs and workloads, ranging between 1k and 1.3M lines of code.
We use the latest available version (1.1.9) and run all benchmarks in the ``speed'' category using the reference input set.
We omit a single benchmark, cam4, as it did not successfully run on our platform.
Table~\ref{table:spec-descriptions} briefly describes each benchmark.

\textbf{DaCapo:}
DaCapo~\cite{10.1145/3669940.3707217} is a widely used benchmark suite for Java applications.
The suite consists of large real-world programs and workloads, ranging between 25k and 6M lines of code.
We use the latest available version (Chopin) and run all benchmarks using the default input set.
Table~\ref{table:dacapo-descriptions} provides a brief description of each benchmark.

\section{Evaluation}
\label{section:eval}
This section addresses the research questions posed in Section~\ref{section:research-questions}, using the causal model and
enhanced methodology introduced in Sections~\ref{section:model} and~\ref{section:methodology}.

\begin{table}[tp]
    \caption{
        \textbf{Versions of compilers and runtime systems used (\textsection\ref{section:eval}).}
        All are most recent as of September 2024.}
    \label{table:versions}
    \centering\begin{tabular}{l | l | l}
        \multicolumn{1}{l|}{\textit{Implementation}} & \multicolumn{1}{l|}{\textit{Language(s)}} & \multicolumn{1}{l}{\textit{Version}}\\ \hline %
        LLVM / Clang & C, C++                 & 19\\
        Rust         & Rust                   & 1.81.0\\
        OpenJDK      & Java                   & 21.0.4+7\\
        Go           & Go                     & 1.23.1\\
        C\# / .NET   & C\#                    & 8.0.8\\
        Node.js      & JavaScript, TypeScript & 20.17.0\\
        PHP          & PHP                    & 8.3.11\\
        tsc          & TypeScript             & 5.6.2\\
        CPython      & Python                 & 3.12.6\\
        PyPy         & Python                 & 7.3.17\\
        Lua          & Lua                    & 5.4.7\\
        LuaJIT       & Lua                    & 87ae18a
    \end{tabular}
\end{table}

We conduct experiments in this paper using a server equipped with two Intel
Xeon Gold 6430 processors totaling 128 logical cores and 128GB of memory, running Linux version
6.8.0-45-generic. Table~\ref{table:versions} details specific compiler and runtime system versions.
Benchmark runs are isolated in Docker containers to ensure reproducibility.
Docker introduces negligible performance overhead~\cite{DBLP:conf/ispass/FelterFRR15},
which we confirm for the experiments detailed in this paper by obtaining equivalent results without Docker.
Docker provides easy access to control groups, making it possible to place
restrictions on CPU usage or to pin programs to execute on specific cores (Section~\ref{section:eval-contributors}).

\subsection{\textbf{RQ1}: Do Some Language Implementations Consume More Energy Than Others?}
\label{section:eval-energy}

\begin{figure}[tp]
    \centering
    \begin{subfigure}{\linewidth}
        \centering
        \includegraphics[width=0.9\linewidth]{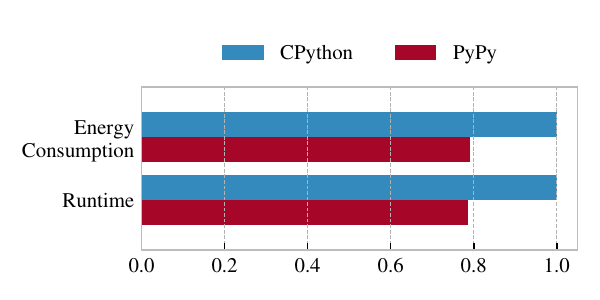}
        \caption{CPython versus PyPy}
    \end{subfigure}
    \begin{subfigure}{\linewidth}
        \centering
        \includegraphics[width=0.9\linewidth]{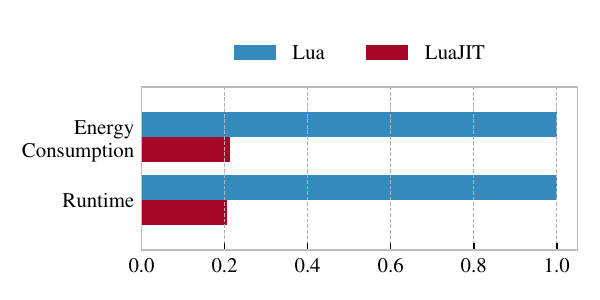}
        \caption{Lua versus LuaJIT}
    \end{subfigure}
    \caption{\textbf{Programming Languages can exhibit different performance characteristics depending on the implementation used (\textsection\ref{section:model-pl-implementation}).}
        Here, we compare interpreters (CPython, Lua) to their JIT equivalent (PyPy, LuaJIT) on the same CLBG benchmark sources.
        We normalize to the interpreter results.
        We find around $5\times$ and $1.25\times$ decreased execution time and energy consumption on average for
        LuaJIT over Lua and PyPy over CPython, respectively.}
    \label{fig:interprets-jits}
\end{figure}

According to our model, the only significant causal path from programming language to energy consumption is mediated by
three factors: the language's implementation, the application's implementation, and execution time.
Therefore, for identical application implementations, power draw variance can only be due to language implementation differences.
Similarly, if the number of cores used is fixed, variance in energy consumption is uniquely determined by execution time.

\textbf{For the Same Language:}
It is well known that different implementations of the same language may have widely varying performance characteristics.
Figure~\ref{fig:interprets-jits} compares languages across multiple implementations on the same set of benchmarks, namely CPython/PyPy and Lua/LuaJIT, and
confirms that not only do the JIT implementations offer increased performance, they also provide corresponding energy savings.
On average for these benchmarks, PyPy is about $1.25\times$ faster than CPython, and LuaJIT is $5\times$ faster than the Lua interpreter.

Note that different implementations of the same language, as well as different versions of the same implementation, may have different tooling, compatibility, or support.
These differences do not affect our experiments.
In our PyPy/CPython and Lua/LuaJIT evaluation, all benchmarks run successfully in both the interpreter and JIT implementation, with only one
exception for each: pidigits does not work with PyPy due to a third-party library compilation failure, and fasta fails in LuaJIT due
to a standard library utility rename.

\textbf{Across Many Languages and Implementations:}
To test if the choice of programming language implementation has an effect on power draw beyond execution time,
we fix other factors that may affect power draw: we limit the program to a single core via Docker's \texttt{-{}-cpuset-cpus} argument,
and pin CPUs to their minimum frequency with turbo mode disabled to ensure frequency scaling and throttling do not increase power draw variance in our measurements.
Once these factors are controlled for, we measure execution time and energy consumption of each benchmark across all language implementations.
These measurements yield equal power draw: $189.8 \pm 0.5$ W. Hence, the impact of choice of programming language implementation (or programming language more generally) on energy consumption beyond execution time is negligible.

\begin{tcolorbox}[colback=green!5!white,colframe=green!75!black, arc=1mm, top=0.5mm, bottom=0.5mm, left=0.5mm, right=0.5mm, enlarge top initially by=1mm]
\textbf{RQ1 Summary:}
Different language implementations can have widely varying performance characteristics, and by extension, energy efficiency.
However, once accounting for external factors such as numbers of cores used, all benchmarks in all language implementations exhibit constant power draw.
Hence, energy consumption is directly proportional to execution time.
Faster language implementations will offer commensurate energy savings.
\end{tcolorbox}

\subsection{\textbf{RQ2}: What Are the Key Contributors to Power Draw Standing Between Choice of Programming Language and Energy Consumption?}
\label{section:eval-contributors}

\begin{figure*}[tp]
    \centering
    \hspace{\fill}
    \begin{subfigure}{0.49\textwidth}
        \centering
        \includegraphics[width=\linewidth]{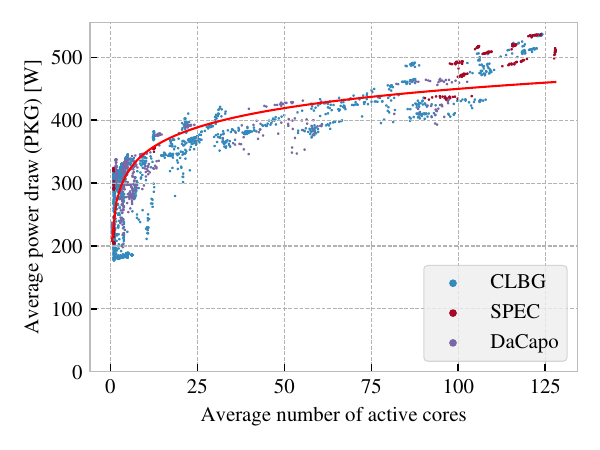}
        \caption{\textbf{Average package (PKG) power draw by average number of active cores.}
            Regression: $y = 31 \log_2 x + 246 \left(R^2 = 0.72\right)$.}
        \label{fig:power-by-cores}
    \end{subfigure}
    \hspace{\fill}
    \begin{subfigure}{0.49\textwidth}
        \centering
        \includegraphics[width=\linewidth]{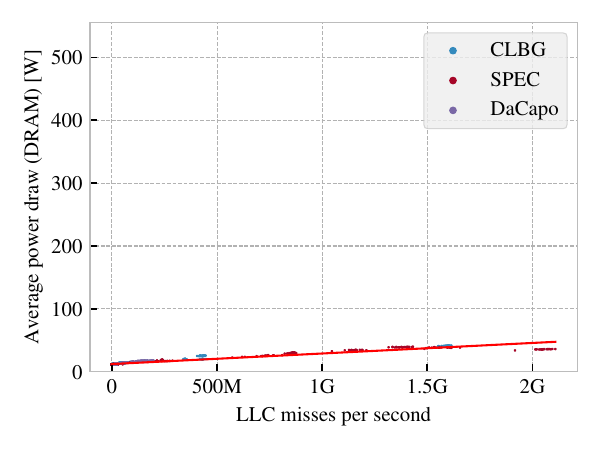}
        \caption{\textbf{Average DRAM power draw by memory activity},
        approximated using LLC misses. Regression: $y = 1.68 \cdot 10^{-8} x + 12 \left(R^2 = 0.93\right)$.}
        \label{fig:memory-activity}
    \end{subfigure}
    \hspace{\fill}
    \caption{\textbf{Power draw is much more significantly affected by the number of active cores than memory activity (\textsection\ref{section:model-power}).}
        Each point on those graphs represents a single benchmark run. }
    \label{fig:power-memory}
\end{figure*}

Our causal model identifies number of active cores and memory activity as primary contributors to power draw.
Section~\ref{section:methodology} details our improved measurement methodology, which allows us to accurately quantify
to which extent these factors actually contribute to power draw.

\textbf{Number of Active Cores:}
Figure~\ref{fig:power-by-cores} shows the relationship between the number of active cores and power draw.
Using multiple cores, energy usage grows up to around $2\times$ the base power draw compared to using a single core.
We
fit a log curve to the data $\left(R^2 = 0.72\right)$, as the relationship does not appear linear, but results may vary across platforms.
Denoting $\mbox{Power}\left(x\right)$ as the average power draw using $x$ cores, we find that, on our system, doubling the number of cores used
increases average power draw by only roughly 31W:
\begin{align*}
    \mbox{Power}\left(x\right) & = 31 \log_2 x + 246\\
    \Rightarrow \mbox{Power}\left(2x\right) & = 31 \log_2 \left(2x\right) + 246\\
    \Rightarrow \mbox{Power}\left(2x\right) & = 31 \log_2 2 + 31 \log_2 x + 246\\
    \Rightarrow \mbox{Power}\left(2x\right) & = \mbox{Power}\left(x\right) + 31
\end{align*}
Further, on our machine, doubling the number of cores increases relative energy efficiency, provided throughput increases by at least 13\%:
\[ \frac{\mbox{Power}\left(2x\right)}{\mbox{Power}\left(x\right)} =
        \frac{31 \log_2 \left(2x\right) + 246}{31 \log_2 x + 246}
        \leq 113\% \quad \left( 1 \leq x \leq 128 \right) \]
In other words, parallelization overhead up to 87\% is acceptable. Therefore, on our experimental platform, aggressively parallelizing programs is nearly always an energy-efficient choice.

\textbf{Memory Activity:}
Previous research has established that roughly 40\% of dynamic random access memory (DRAM) energy consumption is constant
and required to refresh memory cells to keep the system running properly, while the remaining 60\% is related to
read and write activity~\cite{DBLP:conf/isca/SeolSJCSK16}.
Section~\ref{section:measurement-tool} introduces last level cache (LLC) misses as a better proxy for memory activity: any read or write request that cannot
be fulfilled by the LLC will have to go to DRAM. Figure~\ref{fig:memory-activity} shows a strong linear correlation between LLC misses
and increased power draw $\left(R^2 = 0.93\right)$.
However, this figure also highlights how much smaller the memory's energy consumption is to the processor's, with their ratio remaining from 2 to 8 \% across all benchmarks.

\begin{tcolorbox}[colback=green!5!white,colframe=green!75!black, arc=1mm, top=0.5mm, bottom=0.5mm, left=0.5mm, right=0.5mm, enlarge top initially by=1mm]
\textbf{RQ2 Summary:}
Number of active cores is the primary factor in increased power draw. Memory energy consumption is linearly related to
memory usage, but significantly less important than processor energy consumption.
Moreover, aggressive parallelization, even with high overheads, increases energy efficiency.
\end{tcolorbox}

\subsection{\textbf{RQ3}: Can Observed Anomalies in Prior Work Be Explained Through the Lens of Our Causal Model?}
\label{section:eval-examination}

This section identifies four major sources of anomalies in prior work.
For each anomaly, we indicate in brackets which causal path(s) in our model are responsible.

\textbf{Concurrency} [Application Implementation \textrightarrow{} \{Number of Active Cores, Time\}]:
The number of cores used by an application is a major factor in energy consumption, as Section~\ref{section:eval-contributors} quantifies.
Therefore, comparing benchmark implementations in different languages that use different numbers of cores is not a fair comparison.
Level of parallelism imbalances are the main reason for the reported performance discrepancies between JavaScript and TypeScript.
The TypeScript version of the mandelbrot benchmark is $21 \times$ slower than its JavaScript implementation.
Its execution is fully sequential, while the JavaScript version uses 28 cores on average.

Further, since TypeScript is a strict superset of JavaScript and we test them using the same runtime system,
there should be no differences in execution time or energy consumption.
With some minor edits in four benchmarks, all JavaScript benchmark implementations pass the TypeScript compiler without any errors,
and yield equal performance.

\textbf{Third-Party Library Usage} [Application Implementation \textrightarrow{} \{Time, Number of Active Cores\}]:
Choice of third-party libraries used has a significant impact on application performance.
In fact, regular expression benchmarks are often poor candidates to make any comparison beyond the library used.
This is the case for the CLBG regex-redux benchmark: it is $8.9\times$ slower in its C++ implementation compared to the C version.
This difference is entirely due to choice of third-party library: the C version uses the PCRE library,
while the C++ version uses the Boost library.
On this benchmark, Boost's library performs significantly worse than PCRE. This outlier alone accounts
for the entire reported gap between C and C++.

Unlike TypeScript and JavaScript, C++ is not a strict superset of C.
Nonetheless, all C benchmarks compile in C++ mode, only sometimes requiring
the \texttt{-fpermissive} compiler flag or minor changes such as added types or casting.
These programs then yield identical performance and energy consumption numbers.

\textit{Crossing Language Boundaries:}
Using third-party libraries also allows for easily crossing language boundaries, for instance to access
lower-level system utilities or to obtain greater performance. In Python, using NumPy can yield
up to $60,000\times$ faster code when multiplying matrices~\cite{doi:10.1126/science.aam9744, DBLP:conf/osdi/BergerSP23}.
On a small micro-benchmark multiplying $10,000 \times 10,000$ matrices,
we find that NumPy is $84\times$ faster than a naive three-loop C++ implementation,
with $44\times$ lower energy consumption.
Of course, this is because NumPy's underlying matrix-multiply implementation is
written in a low-level language and leverages many optimizations such as 
parallelism, vector instructions, blocking, or hardware-specific instructions.
To some degree, using NumPy and other third-party libraries has become the
``correct'' way of writing Python code: leveraging low-level languages for compute-intensive tasks,
while keeping Python's simplicity.
Even C++ may cross over to code originally written using C, Fortran, or direct Assembly for performance when using BLAS.

\textbf{Just-in-Time (JIT) Warmup} [Programming Language Implementation \textrightarrow{} Time]:
JIT program execution is typically split into a startup phase and a
delayed steady state phase of peak performance. As the program executes, the JIT runtime determines which
code paths are frequently executed, and compiles them into machine code, eventually optimizing hotspots.
However, existing research has already shown that a steady state of peak performance is not
always reached~\cite{DBLP:journals/pacmpl/BarrettBKMT17, DBLP:journals/ese/TrainiCPT23}.
In fact, one of these studies~\cite{DBLP:journals/pacmpl/BarrettBKMT17} uses the same CLBG benchmark suite
and demonstrates that for some benchmarks, a steady state is never reached.

\begin{figure}[tp]
    \centering
    \begin{subfigure}{0.47\textwidth}
        \includegraphics[width=\linewidth]{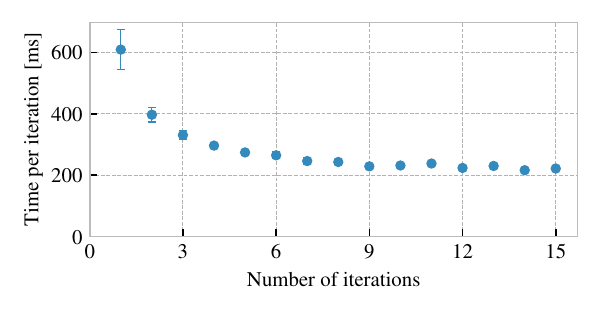}
        \caption{mandelbrot}
    \end{subfigure}
    \hfill
    \begin{subfigure}{0.47\textwidth}
        \includegraphics[width=\linewidth]{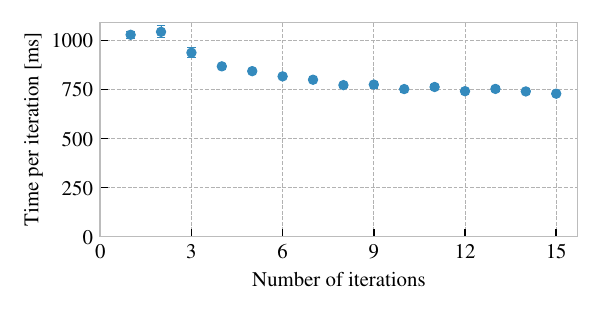}
        \caption{fannkuch-redux}
    \end{subfigure}

    \caption{\textbf{Measuring just the first iteration is not indicative of Java's (OpenJDK) performance for long-lived applications (\textsection\ref{section:model-benchmarks}).}
    We gradually increase the number of iterations, observing decreased per-iteration execution time on several benchmarks, including (a) mandelbrot and (b) fannkuch-redux.}
    \label{fig:java-n}
\end{figure}

Factors other than JIT warmup may also affect the first iteration of short-lived
benchmarks, such as bytecode interpretation or memory and garbage collection initialization. To
verify this, we wrap Java benchmarks in a loop and gradually increase the number of in-process
iterations. Figure~\ref{fig:java-n} shows that for the mandelbrot and fannkuch-redux benchmarks, the first
iteration's execution time is respectively $2.7\times$ and $1.4\times$ slower than when averaging over 15 iterations.
Generally, we see that for these and other benchmarks, averaging over ten or more iterations
amortizes the first iteration's impact: across all CLBG benchmarks,
we observe a mean of 80\% improved energy efficiency and execution time.

\textbf{Garbage Collection} [Programming Language Implementation \textrightarrow{} \{Time, Memory Activity\}]:
An important property of higher-level languages and their implementations is garbage collection, which imposes significant performance overhead~\cite{10.1145/1094811.1094836}.
Out of the 13 language implementations discussed in this paper, 10 use a garbage collector.
Garbage collectors can typically be configured
and tuned for each application to offer maximal performance, which is standard practice in
real-world applications to optimize for application throughput and/or latency.
For example, Go's garbage collector can be disabled
entirely with the \texttt{GOGC} environment variable. When doing so, we observe a $2.8\times$ speedup on the allocation-intensive
binary-trees benchmark, and a $1.23\times$ speedup on mandelbrot.
Disabling the garbage collector over all Go benchmarks
yields an average of $1.15\times$ improved execution time and energy consumption.

\begin{tcolorbox}[colback=green!5!white,colframe=green!75!black, arc=1mm, top=0.5mm, bottom=0.5mm, left=0.5mm, right=0.5mm, enlarge top initially by=1mm]
\textbf{RQ3 Summary:}
Variations in application implementations, notably parallelism and third-party library usage, explain observed anomalies between JavaScript/TypeScript and C/C++.
Important language implementation specifics such as warmup and garbage collection must also adequately be accounted for: when doing so, the unexpectedly large gap between Go/Java and C significantly shrinks.
\end{tcolorbox}

\section{Threats to Validity}
This study largely focuses on reproducing Pereira et al.'s experiments, which rely on a specific set of benchmarks.
We mitigate this threat to external validity by not only focusing on general run-time properties of those programs
rather than their absolute execution time,
but also by verifying results across two additional benchmark suites.

Experiments in this paper are conducted on a server architecture, maintaining comparability with prior work.
While we expect our conclusions to generalize to other architectures, results may differ on mobile, embedded systems, or architectures
mixing energy-saving and high-performance cores.

\section{Other Related Work}
Previous sections, notably Section~\ref{section:prior-work}, discuss and critique the main
line of research relating programming languages and energy consumption~\cite{
10.1145/3125374.3125382, 10.1145/3136014.3136031, DBLP:journals/scp/PereiraCRRCFS21}.
Follow-on studies~\cite{DBLP:journals/sqj/GordilloCMGFAS24, DBLP:conf/ict4s/KoedijkO22}
use external energy measurements rather than on-chip measurements, but still suffer from the same limitations as the original work.

Georgiou et al.~\cite{DBLP:conf/msr/GeorgiouKLS18} explore the Energy Delay Product (EDP) of programming languages on Rosetta Code.
The goal of EDP is to establish a trade-off between energy consumption and execution time.
Our evaluation suggests that, in fact, optimizing for execution time also provides proportional energy consumption reduction.
Georgiou et al.'s public data reveals that this conclusion is in line with their findings:
even in the ``optimize for energy'' case, nearly all benchmark versions which are the fastest are also the most energy-efficient,
with the rare exceptions attributable to short execution times, coarse measurement granularity, and/or missing multicore accounting.
Our methodology leverages modern on-chip energy and performance counters, offering an analysis and new causal model based on more granular and precise measurements.

Several other studies~\cite{DBLP:conf/usenix/LionCSY22, DBLP:conf/icse/NanzF15} explore and compare performance and other properties across programming languages, but do not address energy efficiency directly.
These comparative studies are complementary to this work: we show here that execution time is the primary factor in energy consumption.

Processor operating voltage and frequency have an impact on energy consumption.
This impact has been studied in the context of trading off speed for reduced energy consumption,
where compiling for speed yields better energy
efficiency in the general case~\cite{DBLP:conf/lcpc/YukiR13}.
For specific applications, such as sparse matrix computations,
compile-time techniques may
decrease energy usage without impacting execution time by leveraging
load imbalance~\cite{DBLP:conf/ipps/ChenMKR05}.
Other work has shown that using characteristics of a program during execution
to reduce the processor's voltage or frequency can yield higher energy efficiency for a comparably
smaller performance degradation~\cite{DBLP:conf/sc/HsuF05}.
For example, performance counter information can be used to scale
the processor's frequency~\cite{10.1145/581630.581668}.
These approaches are orthogonal to the choice of programming language and could be applied in many
programming environments.

Graphics processing units (GPUs) are increasingly prevalent in modern computing systems, notably
in artificial intelligence workloads.
Past work has examined the energy efficiency of GPUs from a software perspective,
notably accounting for the energy expenditure of tensors~\cite{10.1145/3597503.3639156}
and comparing the energy consumption of different machine learning frameworks~\cite{10.1145/3510003.3510221}.
Extending our work to incorporate a model for energy consumption of GPUs is a potential avenue for future work.

Previous research also addresses the energy consumption of different
data structures in certain languages, notably Java
collections~\cite{10.1145/2884781.2884869, DBLP:conf/greens/PereiraCSCF16, DBLP:journals/ese/OliveiraOCPF21},
or Python data manipulation libraries~\cite{10.1145/3661167.3661203}.
Data structures can be interesting candidates
to study through an energy-focused lens as they can have a substantial impact on locality and memory
activity, which, as Section~\ref{section:model-power} describes, are factors in energy consumption.
However, programmers often have a limited choice of
data structures that fit their use case or algorithm.

Finally, past work has also investigated energy efficiency of concurrent programs, focusing on
Haskell's data sharing primitives~\cite{DBLP:conf/wcre/LimaSLCMF16} or Java's thread management constructs~\cite{10.1145/2660193.2660235},
or analyzing the power to performance trade-offs in lock-based synchronization~\cite{DBLP:conf/usenix/FalsafiGPT16}.
As Section~\ref{section:model-power} argues, these studies confirm that the power usage patterns of the processor
and the machine as a whole are more complex with parallel execution.

\section{Conclusion}
This paper presents a detailed causal model exploring the complex relationship
between choice of programming language and energy consumption. It captures
some of the factors at play, notably distinguishing implementation
from programming language, and establishing the number of active cores and memory activity
as important factors in power draw.

Using this causal model, we investigate and explain anomalies in previous research,
finding that many factors such as parallelism level, benchmark implementation specifics,
or language implementation properties must be taken into account for a fair comparison.
Our results suggest that the
choice of programming language has no significant impact on energy consumption
beyond execution time. Programmers aiming to reduce energy consumption can do so by focusing on performance optimizations.
This strategy is possible even in
``inefficient'' programming languages like Python by using faster language implementations,
employing faster and more parallel algorithms, and using native libraries.

\section*{Data-Availability Statement}
Data supporting this paper's evaluation is available at
\texttt{\href{https://github.com/nicovank/Energy-Languages}{github.com/nicovank/Energy-Languages}}.

\section*{Acknowledgments}
We thank Prashant Shenoy, Walid Hanafy, Noman Bashir, Mehmet Savasci, and Abel Souza
for their help getting started with energy measurements using RAPL, their
continuous feedback, and for providing access to an energy-measurement-ready machine.
We thank Eunice Jun for suggesting exploration of causal models in this work.

\bibliographystyle{IEEEtran}
\bibliography{IEEEabrv,references}{}
\end{document}